\documentclass[12pt]{article}

\title{Cube or hypercube of natural units
\thanks{To be published in ``Multiple facets of quantization and
supersymmetry'', Michael Marinov Memorial Volume, Eds. M.
Olshanetsky and A. Vainshtein, World Scientific, 2002.} }
\author{L.B. Okun \\
ITEP, Moscow, 117218, Russia \\
email: okun@heron.itep.ru}
\date{}

\begin{document}
\maketitle

\begin{abstract}

Max Planck introduced four natural units: $h, c, G, k$. Only the
first three of them retained their status, representing the so
called cube of theories, after the theory of relativity and
quantum mechanics were created and became the pillars of physics.
This short note is a little pebble on the tombstone of Michael
Samuilovich Marinov.
\end{abstract}

\newpage

\section*{Dedication}

In the evening of January 20, 2000 an email from ITEP was sent to
Serezha Gurvitz and Arnon Dar:
\begin{quote}
``Dear Serezha,

        ~\,Dear Arnon,

It is with great sorrow that we learned that Misha passed away. He
was a brilliant physicist, an extraordinary person of outstanding
integrity and courage, a friend of great warmth. He was like a
tuning fork for all of us when he was in Moscow and later in
Israel. We recall his lectures on group theory at ITEP, we recall
his remarkable lectures on path integral. We recall his deep and
original papers. He was a source of great knowledge and great
wisdom.

Today, on the Monday seminar, there was a minute of silence to the
memory of Mikhail Samuilovich Marinov.

Please, forward our condolences to Lilya, Masha, and Dina.

Lev Okun, Victor Novikov, Mikhail Vysotsky''.
\end{quote}


\noindent
Today, two years later, I subscribe to every word written that
evening. This short note is a little pebble on the tombstone.

\section{Planck units}

Max Planck during years 1897--1899 published in the
Sitz.-Ber.\ Preuss.\ Akad.\ Wiss.\ five consecutive reports\,\cite{1}
with the same title
``\"{U}ber irreversible Strahlung\-svorg\"{a}nge'' [On irreversible
 processes of radiation]. Under the same title the article\,\cite{2}
 appeared in Annalen der
Physik in 1900, which summarized his reports\,\cite{1} and his talk
in Munich. In the article {\bf (e)} in ref.\cite{1} and in ref.\cite{2}
 a special
section was added on natural units of measure, in which Planck wrote (in an
abridged translation from German):

\begin{quote}
``All systems of physical units including the so-called absolute
C.G.S.-system, appeared up to now due to accidental circumstances,
as the choice of basic units in each of these systems occurred not
from a general point of view valid for any place and time, but
from the needs of our earthly culture. In this connection it would
be interesting to note, that by using both constants $a$ and $b$,
which appear in the equation (41) for the entropy of radiation we
get the possibility to establish units of length, mass, time and
temperature, which would not depend on the choice of special
bodies or substances and would be valid for all epochs and all
cultures including extraterrestrial and extrahuman ones and could
therefore serve as `natural units of measurements' ''.\cite{2}
\end{quote}

Let us note that equation (41) in ref.\cite{2} had the form: $$ S=
-\frac{U}{a\nu} \,\ln\frac{U}{eb\nu} \;\; , $$ where $S$ is the
total electromagnetic entropy, $U$ is the total energy of the
system, $\nu$ is the frequency of radiation, $e$ is the base of
the system of natural logarithms, $a=h/k$, $b=h$ in modern
notations, where $h$ is the Planck constant, while $k$ is Boltzman
constant.

Already in 1900 Planck wrote his famous formula for the spectral
energy density: $$ \frac{8\pi\nu^3
h}{c^3}\frac{d\nu}{e^{h\nu/kT}-1} $$ using notations $h$ and $k$
instead of $b$ and $b/a$ respectively.\cite{3} The
references\,\cite{1,2,3} have been reprinted in ref. \cite{4}.

\begin{quote}
``\ldots The four units -- for length, mass, time and temperature
-- are expressed in terms of above mentioned constants $a$ and
$b$, velocity of light $c$, and gravitational constant $f$. When
expressed in centimeters, grams, seconds and Celsius degrees,
these four quantities have the following values: $$
\begin{array}{ll}
a~ = & 0.4818 \cdot 10^{-10}\; {\rm sec} \cdot~\!\! ^0{\rm C} \\[1mm]
b~ = & 6.885 \cdot 10^{-27}\; {\rm cm}^2 \cdot {\rm g}/{\rm sec} \\[1mm]
c~ = & 3.00 \cdot 10^{10}\; {\rm cm}/{\rm sec} \\[1mm]
 f~ = & 6.885 \cdot 10^{-8}\; {\rm cm}^2/{\rm g} \cdot {\rm sec}^2 \;
\mbox{\rm \ldots ''
\cite{2}}
\end{array}
$$
\end{quote}

Note that the modern notations and values are (see Review of Particle
Physics \cite{5}):
$$
\begin{array}{ll}
h/k~(= a)~ = & 0.4799237 \cdot 10^{-10} \; {\rm s}\cdot {\rm K} \\[1mm]
k~(= b/a)~ = & 1.3806503(24) \cdot 10^{-23}\; {\rm J}
\cdot {\rm K}^{-1} \\[1mm]
h~ (= b~)~~~ = & 6.62606876(52) \cdot 10^{-34}\; {\rm J} \cdot {\rm s} \\[1mm]
c \qquad\qquad= & 299792458\; {\rm m\; s}^{-1} \\[1mm]
G~(=f~)~~ = & 6.673(10) \cdot 10^{-11}\; {\rm m}^3\, {\rm kg}^{-1}\, {\rm s}^{-2}
\; ,
\end{array}
$$ where K  is Kelvin degree ($1 {\rm K} = 1~^0{\rm C}$).
\begin{quote}
``If we  chose now the ``natural units'', so that each of the
above constants is put equal to 1, then we will get
$$
\begin{array}{lll}
\mbox{for the unit of length:} & \mbox{\large$\sqrt{\frac{b f}{c^3}}
\left(=\!\sqrt{\frac{hG}{c^3}}\right) \quad=$} & 4.13 \cdot 10^{-33} \; {\rm cm}
\; , \\[2mm]
\mbox{for the unit of mass:} & \mbox{\large$\sqrt{\frac{b c}{f}}
\left(=\!\sqrt{\frac{hc}{G}}\right)$} \quad ~~= & 5.56 \cdot 10^{-5} \; {\rm g} \;
, \\[2mm]
\mbox{for the unit of time:} & \mbox{\large$\sqrt{\frac{b f}{c^5}}
\left(=\!\sqrt{\frac{hG}{c^5}}\right)$} \quad ~= & 1.38 \cdot 10^{-43} \; {\rm s}
\;,
\\[2mm]
\mbox{for the unit of temperature:} & a\mbox{\large$ \sqrt{\frac{c^5}{b f}}
\left(=\!\frac{1}{k}\sqrt{\frac{hc^5}{G}}\right)$}\!= & 3.50 \cdot 10^{32} \;
^0{\rm C}
\; .
\end{array}
$$

These units will have their natural meaning as long as the laws of
gravitation and of light propagation as well as both principles of
thermodynamics remain valid.'' \cite{2}
\end{quote}

Thus Planck considered four natural constants. This was natural for
him because the expression $e^{h\nu/kT}$ in the formula for energy
density contains both $h$ and $k$ on equal footing and because
quantum mechanics and relativity were unknown at that time, while
principles of thermodynamics were considered to be fundamental.

\section{The cube of theories}

It is interesting that at the beginning of the article by G.
Gamov, D. Ivanenko and L. Landau\,\cite{6} they refer to the
natural Planck system of four units, while in the main body of
this article they consider as natural only three Planck units
without $k$ and temperature. This approach had been taken up by
M.~Bronshtein\,\cite{7,8} and Zelmanov,\cite{9,10} who developed
the idea of the cube of theories and by their
followers.\cite{11,12}

The cube is located along three orthogonal axes marked by $c$
(actually by $1/c$), $\hbar$, $G$. The vertex ($000$) corresponds
to nonrelativistic mechanics,($c00$) -- to special relativity,
($0\hbar0$) -- to non-relativistic quantum mechanics, ($c\hbar0$)
-- to quantum field theory, ($c$0$G$) -- to general relativity,
($c$$\hbar$$G$) -- to futuristic quantum gravity  and the Theory
of Everything, TOE. There is a hope that in the framework of TOE
the values of dimensionless fundamental parameters  will be
ultimately calculated.

\section{Temperature and Entropy}

Let us note that in ``Statistical Physics'' by L.~Landau and
E.~Lifshitz\,\cite{13} temperature is measured in units of energy
and hence in all formulas $kT$ is substituted by $T$ ($k$ is put
to 1). Mathematically temperature $T$ is defined as a derivative
of internal energy $U$ of a system over its entropy $S$: $$ T=
dU/dS \;\; . $$

By keeping $k$ Planck contradicted his own definition of natural
units, because $k$ is a ratio of ``hand-crafted'' units of
temperature and energy. As temperature is an average energy of an
ensemble of particles, it is natural to measure it in units of
energy. In fact $k$ is not a natural unit, but a conversion factor
from degrees to joules or electronvolts. Therefore one should
consider a three-dimensional cube of natural units (as was
presented in ref.\cite{6}$^-$\cite{12}$^,$\,\cite{14}), not a four
dimensional hypercube, as was suggested by
J.~Fr\"{o}hlich\,\cite{15}, who following Planck included $k$ into
the set of fundamental units.

The way one treats $k$ is intimately connected with the way one
treats en\-tropy. According to Planck (and to many modern
text-books) entropy looks dimensionful: $[S] = [k]$. The book by
Landau and Lifshitz\,\cite{13}, according to which physical entropy
is dimensionless, is one of the rare exceptions. On the other hand
the informational entropy is usually defined as a dimensionless
quantity:
$$
S_{\rm inf} = -\sum^n_{k=1}P_k \ln P_k \;\; ,
$$
where $P_k$ is probability of finding $x_k$ from a set $x_1, x_2
... x_n$ which describes a message. $S_{\rm inf} =0$, if one of
$P_k =1$, while all other $P_k$ vanish. $S_{\rm inf}$ is maximal when
all $P_k$ are equal (in this case the uncertainty of information
is maximal). Thus in the case of $k=1$ the definitions of physical
and informational entropy are similar.

\section*{Acknowledgements}

I am very grateful to Ya.I. Granowsky and V.L. Telegdi for
interesting remarks. The work was supported by A. von Humbold
Award and grant of RFBR 00-15-96562.

\end{document}